\documentclass[letterpaper,12pt]{article}
\usepackage{url,amsfonts}
\usepackage{array}
\usepackage[pdftex]{graphicx}
\usepackage{subfig}
\usepackage{endfloat}
\usepackage{booktabs}
\usepackage{caption}
\usepackage{indentfirst} 
\usepackage{setspace}
\usepackage{latexsym}
\usepackage{amsmath}
\usepackage{amssymb}
\usepackage{amsthm}
\usepackage{siunitx}
\usepackage[version=3]{mhchem}
\usepackage{braket}
\usepackage{palatino}
\renewcommand{\vec}[1]{\mathbf{#1}}
\usepackage{cite}

\newcommand{\beq}{\begin{equation}}
\newcommand{\eeq}{\end{equation}}
\newcommand{\beqa}{\begin{eqnarray}}
\newcommand{\eeqa}{\end{eqnarray}}
\newcommand{\bfig}{\begin{figure}\begin{center}}
\newcommand{\efig}{\end{center}\end{figure}}
\newcommand{\btab}{\begin{table}\begin{center}}
\newcommand{\etab}{\end{center}\end{table}}
\newcommand{\equ}[1]{Eq.~(\ref{#1})}

\newcommand{\rvec}{\ensuremath{\mathbf{r}}}
\newcommand{\Rvec}{\ensuremath{\mathbf{R}}}

\newcommand{\kvec}{\ensuremath{\mathbf{k}}}

\newcommand{\der}[2]{\frac{\partial #1}{\partial #2}}

\usepackage{fullpage}

\begin{document}
\begin{center}
\vspace*{1cm}
{\LARGE\bf Periodic Subsystem Density-Functional Theory}\\[3ex]

{\large Alessandro Genova,$^1$  Davide Ceresoli,$^{1,2}$ and Michele Pavanello$^1$}\\
$^1$Department of Chemistry, Rutgers University, Newark, NJ 07102, USA\\
$^2$CNR-ISTM: Institute of Molecular Sciences and Technologies, Milano, Italy\\[2ex]
\end{center}

\vfill

\begin{tabbing}
Date:   \quad\= \today \\
Status: \> Submitted to J.\ Chem.\ Phys.\ \\
\end{tabbing}
\newpage

\begin{abstract}
By partitioning the electron density into subsystem contributions, the Frozen Density Embedding (FDE) formulation of subsystem DFT has recently emerged as a powerful tool for reducing the computational scaling of Kohn--Sham DFT. To date, however, FDE has been employed to molecular systems only. Periodic systems, such as metals, semiconductors, and other crystalline solids have been outside the applicability of FDE, mostly because of the lack of a periodic FDE implementation.
To fill this gap, in this work we aim at extening FDE to treat subsystems of molecular and periodic character. This goal is achieved by a dual approach. On one side, the development of a theoretical framework for periodic subsystem DFT. On the other, the realization of the method into a parallel computer code.
We find that periodic FDE is capable of reproducing total electron densities and (to a lesser extent) also interaction energies of molecular systems weakly interacting with metallic surfaces. In the pilot calculations considered, we find that FDE fails in those cases where there is appreciable density overlap between the subsystems. Conversely, we find FDE to be in semiquantitative agreement (but still within chemical accuracy) with Kohn--Sham DFT when the inter-subsystem density overlap is low.
We also conclude that to make FDE a suitable method for describing molecular adsorption at surfaces, kinetic energy density functionals that go beyond the GGA level must be employed.
\end{abstract}

\newpage

\section{Introduction}
\label{sec:intro}

An important theoretical simplification to electronic structure theory was provided by the Hohemberg and Kohn theorems \cite{hohe1964}, according to which either the electron density, $\rho(\rvec)$, or the electronic wavefunction, can be considered as the central quantity in an electronic structure calculation. In other words, $\rho(\rvec)$ provides all information one needs to evaluate the electronic energy and observables of a molecular system. In practical calculations $\rho(\rvec)$ is computed with the Kohn--Sham DFT (KS-DFT) method \cite{kohn1965}, which has by now been implemented in a multitude of computer codes worldwide. KS-DFT is nowadays the most popular quantum mechanical method for calculating the electronic structure of molecules and materials.

Because of the $N^3$ scaling with the sistem size, KS-DFT is limited to the type of sistems it can approach. Tipically, when theoretically modeling of a physicochemical process, a careful choice of the model system precedes the computations. A model system approachable by KS-DFT computations generally should not exceed 500 atoms, so that the simulations can be completed in a reasonable time. However, it is now understood that if chemical accuracy in the predictions is sought, the KS-DFT simulations must take into account the complexity of the environment surrounding the model system of interest \cite{doum2012,dean2011,bene2013,neug2010}. Thus, the formulation of algorithms that simplify the KS-DFT problem by exploiting the locality of the electronic structure have been a popular avenue of research \cite{bowl2012}. 

A natural way to simplify any problem is by applying the principle of divide and conquer. Such an approach is achieved in KS-DFT by partitioning the very basic quantity of DFT: the electron density. Namely,
\begin{equation}
\label{eq:int1}
\rho (\rvec) = \sum_I^{N_S} \rho_I (\rvec).
\end{equation}

As we explain in more detail in the next section, such a partitioning results into a set of coupled differential equations which solve for each subsystem electron density, $\rho_I$. 

This is not a new idea. For example, the frozen density embedding (FDE) formulation of subsystem DFT introduced by Wesolowski and Warshel \cite{doi:10.1021/j100132a040}, has emerged as the leading approach to exploit \equ{eq:int1}. Since that first formulation, the FDE method has been implemented in several quantum chemistry packages that use atomic orbital basis sets, such as ADF \cite{JCC:JCC20861} and TurboMole \cite{doi:10.1063/1.3494537}.  FDE has made possible the all-electron description of such large systems as pigment aggregates \cite{koni2011}, bulk liquids \cite{Iannuzzi200616}, solvated chromophores \cite{neug2005f,neug2005b}, molecular crystals \cite{kevo2013}, as well as biological systems \cite{jctc10-2546,solo2012,jaco2008} to name a few applications. 

Another appealing feature of \equ{eq:int1}, is that it provides us with a molecule-centric description of nature. As many systems naturally occur as a set of weakly interacting molecules, a subsystem formulation of KS-DFT offers an intuitive description of the underlying physics and an easier interpretation of the chemistry as advocated by the earliest attempts \cite{gord1972,kolos1978}. Such a molecule-centric perspective, for example, has motivated many other types of partitioning, e.g.\ the ones occurring at the wavefunction level \cite{jezi1994,luka2010}.

The possibility of applying FDE to periodic systems is of particular interest. Systems such as metals and semiconductors are involved in a plethora of applications important to chemistry and materials science, ranging from photovoltaics to catalysis. A major motivation of this work resides in the success of early implementations of Wavefunction in periodic DFT embedding schemes \cite{Govind1998129}. The CASTEP implementation \cite{Govind1998129}, was employed in simulations involving two subsystems, of which one was treated at the correlated wavefunction level \cite{huan2006,klue2002,klue2001}. There is another periodic implementation of FDE in the CP2K code. This implementation, however, features a Brillouin zone sampling of the $\Gamma$-point only \cite{Iannuzzi200616} and, to the best of our knowledge, it was only applied to the dynamics of liquid water.

In this work, we build on the mentioned early work, and extend the FDE method to approach a system composed of multiple subsystems (2 or more) in which one or more subsystems can be periodic (e.g.\ metals or semiconductors). For this purpose, we introduce a novel implementation of FDE in the {\sc Quantum ESPRESSO} (QE) \cite{QE-2009} suite of softwares.  QE uses pseudopotentials and expands the so-called pseudowaves and the pseudodensity in a plane waves basis set. This attractive framework allows us to account for the non-vanishing Brillouin zone using a $k$-point mash so that we can model accurately the electronic structure of metal and semiconducting surface slab subsystems \cite{monk1976}.  

There are known weaknesses of the FDE method, which are related to the fact that FDE employs orbital--free Kinetic energy density functionals (KEDF). Unlike the exchange--correlation, semilocal KEDF approximants employed in a pure orbital--free framework are unable to reproduce basic aspects of the electronic structure of atoms, such as the shell structure \cite{wang2000}. In FDE, the KEDF are used only at the non-additive level (vide infra). In this role, they are shown to be successful provided that there is weak overlap between electron densities of the subsystems. Weakly interacting subsystems, such as  hydrogen-bonded systems \cite{kevo2006,jcp105-9182,jcp106-8516,jcp128-044114,jctc5-3161,jacs126-11444}, van der Waals complexes \cite{kevo2006,jcp108-6078,jcp116-6411,jcp123-174104}, ionic bonds, and even charge transfer systems \cite{jctc10-2546,solo2014,solo2012} are shown to be successfully reproduced by FDE.

Due to the lack of an extension of FDE to periodic systems (including $k$-point sampling), the ability of FDE to reproduce those interactions arising between molecular adsorbates  and metallic or semiconducting surfaces has so far not been investigated. In this work, we make significant steps forward to fill this gap, and test a number of non-additive KEDF specifically for molecules at metal surfaces. 

This paper is organized as follows. The following section is completely devoted to the theoretical background of FDE when periodic systems are considered. Section 3 is devoted to the computational details of our calculations, while Section 4 presents numerical results of our FDE implementation applied to molecular and periodic systems. Section 5 collects conclusions and future directions of this work.

\section{Theory of Periodic Frozen Density Embedding}
\label{sec:method}

In this section, we will discuss the theory behind periodic subsystem DFT. As our goal is to tackle periodic systems, the equations derived for practical calculations (which we call Frozen Density Embedding, or FDE) are generalized here to a non-integer subsystem orbital occupation, and to a non-vanishing size of the Brillouin zone. Although we recognize a possible semantical inconsistency \cite{weso1996b}, hereafter, we will consider FDE a synonym of subsystem DFT.

The foundamental idea behind any subsystem DFT method is that the electron density of the supersystem can be partitioned into the electron densities of the $N_S$ subsystems, as already shown in \equ{eq:int1}:
\begin{equation}
 \label{eq:meth:1}
 \rho_\text{tot} (\rvec) \equiv \rho (\rvec) = \sum_I^{N_S} \rho_I (\rvec)
\end{equation}
The subsystem densities are defined from the orbitals of the corresponding subsystem (for the sake of clarity only the closed-shell case is reported here):
\begin{equation}
 \rho_I (\rvec) = \frac{2}{\Omega_\text{BZ}} \int_\text{BZ} \text{d}\kvec ~\sum_{j} n_{j,\kvec}^I\left| u_{j,\kvec}^I(\rvec) \right|^2 
\end{equation}
where $\Omega_{\text{BZ}}$ is the volume of the BZ, and $0 \leq n_{j,\kvec}^I \leq 1$ are the occupation numbers of orbitals belonging to fragment $I$, and $u_{j,\kvec}^I(\rvec)$ is such that the Bloch function is $\psi_{j,\kvec}=e^{i\kvec\cdot\rvec}u_{j,\kvec}^I(\rvec)$. The occupation numbers are formulated such that 
\begin{equation}
\sum_{j}  n_{j,\kvec}^I=N_I ~~ \forall ~\kvec \in \text{BZ},
\end{equation}
with $N_I$ the number of electrons assigned to subsystem $I$. The above definition of subsystem density differs from what has been considered in previous formulations of FDE. Here, besides an integral over the BZ, partial orbital occupations are invoked. There are two reasons for this. First a practical one, when tackling metallic systems, due to the continuous density of states at the Fermi energy, the SCF procedure would simply not converge without smearing the occupations across the Fermi energy. Second, when working in a finite temperature regime, the non-pure state resulting form the statistical occupation of excited states can be formally accounted for with a non-integer orbital occupation \cite{parr1989}.

Since the orbitals of one subsystem are not required to be orthogonal to those in another subsystem, complications in the computation of the total non-interacting kinetic energy arise. We can still define a subsystem non-interacting (Janak) kinetic energy as:
\begin{equation}
 T_\text{J} [\rho_I] = \frac{2}{\Omega_\text{BZ}}  \int_\text{BZ} \text{d}\kvec ~ \sum_{j}  n_{j,\kvec}^I \Braket{u_{j,\kvec}^I|-\frac{\left(\nabla+{i}\kvec\right)^2}{2} |u_{j,\kvec}^I}
\end{equation}
We are adopting this notation rather than $T_s$, because the latter is defined for integer occupations only \cite{parr1989}.

As the supersystem total Janak kinetic energy is not simply the sum of the subsystem kinetic energies, it needs to be corrected. Namely,
\begin{equation}
 T_\text{J}[\rho] = \sum_I T_\text{J} [\rho_I] + T_\text{J}^\text{nad}[\{\rho_I\}]
\end{equation}

The non-additive kinetic energy term is defined as:
\begin{equation}
 T_\text{J}^\text{nad}[\{\rho_I\}] = \tilde{T}_\text{J} [\rho] - \sum_I \tilde{T}_\text{J} [\rho_I]
\end{equation}
where $\tilde{T}_\text{J}$ differs from $T_\text{J}$ in the fact that it is a pure functional of the electron density. This approach is particularly useful in practical implementations of the method as it allows us to avoid the diagonalization step of the Fock matrix of the supersystem, effectively making FDE a linear scaling method.

We can also write other energy contributions in a similar fashion:
\begin{equation}
  F [\rho] = F [\{\rho_I\}] = \sum_I F [\rho_I] + \tilde{F} [\rho] - \sum_I \tilde{F} [\rho_I] = \sum_I F [\rho_I] + F^\text{nad} [\{\rho_I\}],
\end{equation}
where $F[\rho]$ can be any functional of the total electron density, such as the Hartree energy $E_H$ or the exchange--correlation energy $E_{XC}$.

This trick allows us to write the total energy of the supersystem as the KS-DFT energies of the single subsystems, plus a contribution arising from the interaction with the other subsystems:
\begin{equation}
 E_\text{FDE} = \sum_I \left\{ T_\text{J} [\rho_I] + E_{eN}^I [\rho_I] + E_{H} [\rho_I] + E_{XC} [\rho_I] \right\} + \sum_I^{N_S}\sum_{J\neq I}^{N_S}E_{eN}^J[\rho_I] + T_\text{J}^\text{nad} + E_H^\text{nad} + E_{XC}^\text{nad} + V_{NN} 
\end{equation}
where $E_{eN}^J[\rho_I]$ is the electron--nuclear interaction energy due to the nuclei of subsystem $J$ with the electrons of subsystem $I$, namely
\begin{equation}
E_{eN}^J[\rho_I]= - \sum_{\alpha \in J} \int \frac{Z_\alpha\rho_I(\rvec)}{|\rvec - \Rvec_\alpha|} \text{d}\rvec. 
\end{equation}
In addition, $V_{NN}$ is the nuclear repulsion energy.

We obtain the subsystem orbitals by solving self-consistently the following coupled equations: 
\begin{equation}
\label{eq:KS}
 \left[ - \frac{1}{2} \left( \nabla + {i}\kvec \right)^2 + V_\text{eff}^I(\rvec) \right] u_{j,\kvec}^I(\rvec) = \epsilon_{j,\kvec}^I u_{j,\kvec}^I(\rvec).
\end{equation}
In the above equation, as we takle periodic systems the effective Hamiltonian for periodic systems is invoked which includes the $\kvec$ vector belonging to the BZ, solving for the periodic part of the Bloch wave. The effective potential $V_\text{eff}^I$ is given by
\begin{equation}
 V_\text{eff}^I(\rvec) = V_{eN}^I(\rvec) + V_H[\rho_I](\rvec) + V_{XC}[\rho_I](\rvec) + V_\text{emb}^I(\rvec),
\end{equation}
and $V_\text{emb}^I$ appearing above is the so-called embedding potential, which takes the following form
\begin{equation}
\label{eq:embedding}
 V_\text{emb}^I(\rvec) = \sum_{J\neq I}^{N_S} \left[ \int \frac{\rho_J(\rvec')}{|\rvec - \rvec'|} \, d\rvec' - \sum_{\alpha \in J} \frac{Z_\alpha}{|\rvec - \Rvec_\alpha|} \right] + \der{\tilde{T}_\text{J}[\rho]}{\rho} - \der{\tilde{T}_\text{J}[\rho_I]}{\rho_I} + \der{\tilde{E}_{XC}[\rho]}{\rho} - \der{\tilde{E}_{XC}[\rho_I]}{\rho_I}.
\end{equation}

We refer to other review publications \cite{weso2006,jaco2013,grit2013} regarding the more profound theoretical ramifications of using the partitioning in \equ{eq:meth:1}, and the assumptions related to the non-interacting $v$-representability of each subsystem density.
We will limit ourselves in stating that the more accurate the non-additive KEDF is the closer the FDE results converge to KS-DFT of the supersystem \cite{jaco2013}. As the exact functional is unknown, the most important avenue of research in FDE is the systematical improvement of the available KEDF approximants. 

\begin{figure}[p]
 \centering
 \subfloat[][]{\label{fig:mpi}
 \includegraphics[width=.7\columnwidth]{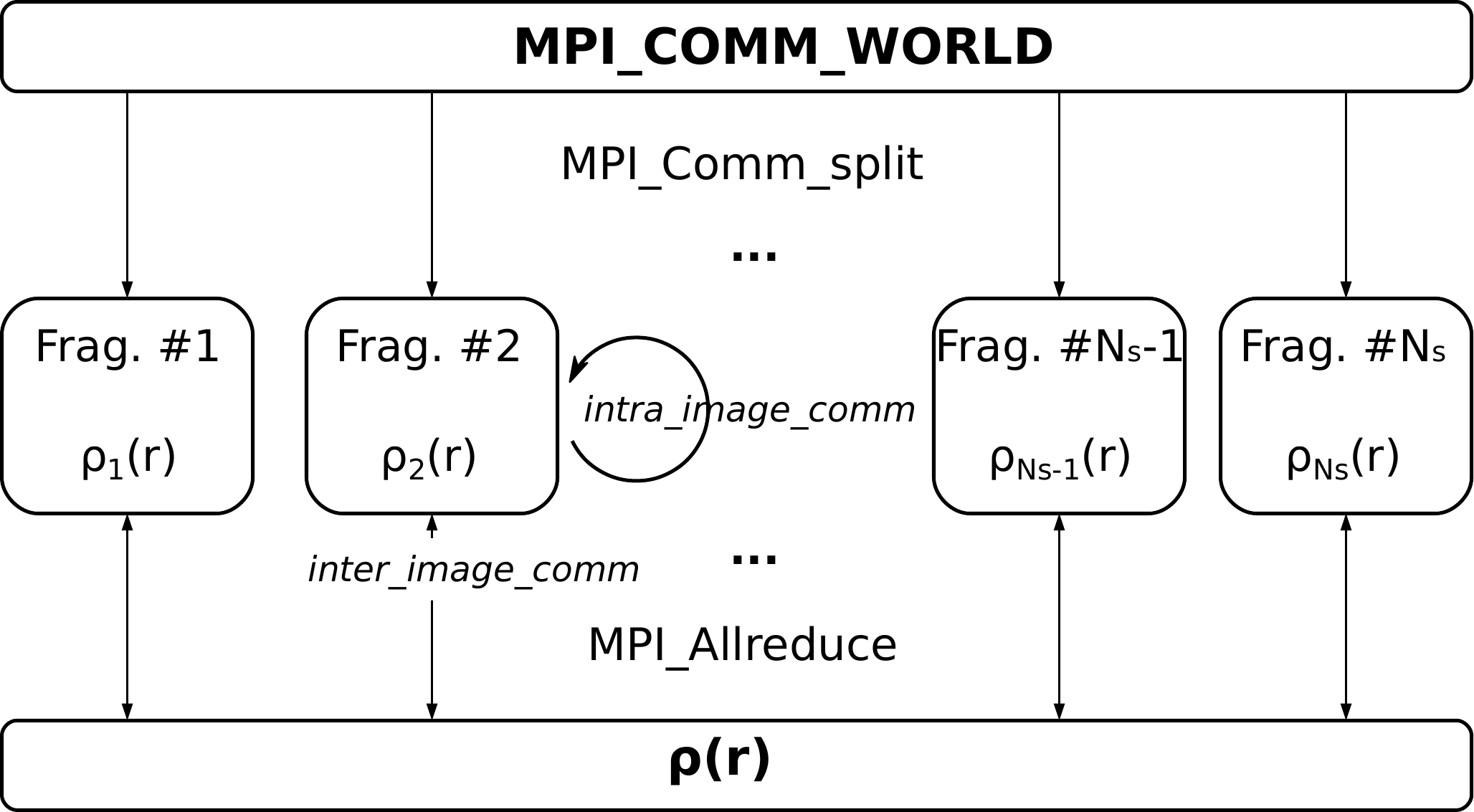}} \\
 \subfloat[][]{\label{fig:SCF}
 \includegraphics[width=.4\columnwidth]{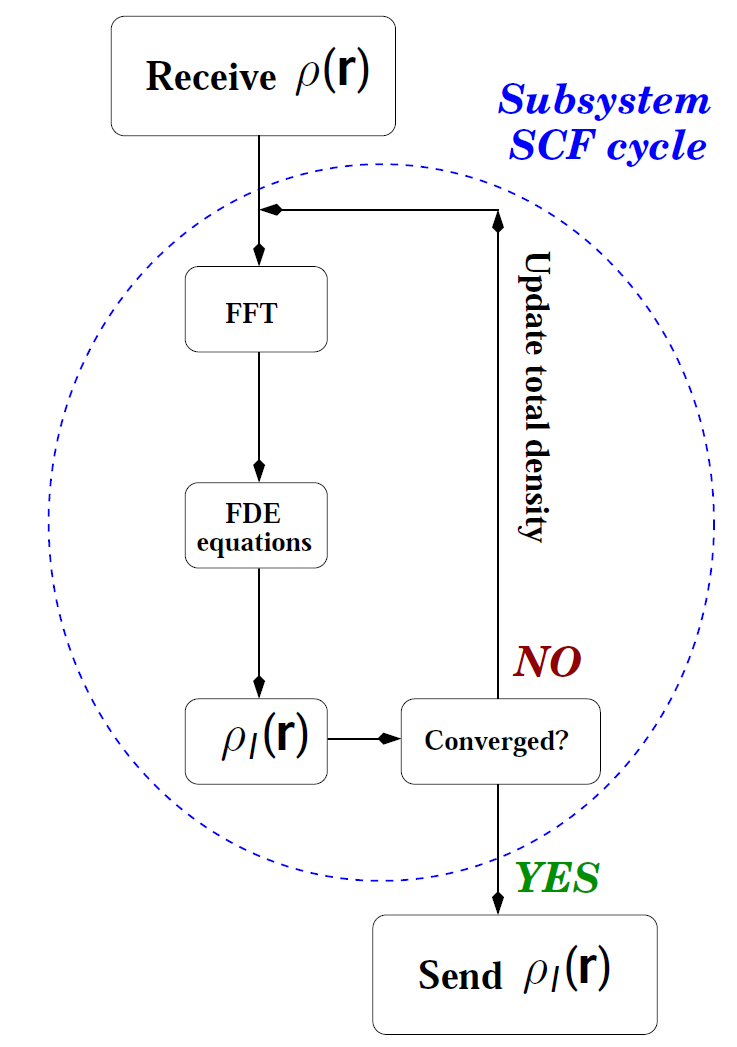}}
   \caption[Parallel architecture]{\subref{fig:mpi}: The structure of the MPI routines. The MPI routines are responsible for ``summing'' all the subsystem densities, $\{\rho_I\},~I=1,2\ldots N_S$ and broadcasting the toal density, $\rho$, to all processes. Subsystems communicate through the inter-image communicator, while the various processes solving for the single subsystem electronic structure communicate with the intra-image communicator. \subref{fig:SCF}: The subsystem SCF cycle. ``FDE equations'' stands for \equ{eq:KS}. If the SCF is executed to selfconsistency before broadcasting the $\{\rho_I\}$, the freeze-and-thaw preocedure is recovered. In our code, we allow the total density to be updated at every SCF cycle so that when the SCF has converged all subsystem densities are computed selfconsistently to each other.}
 \label{fig:parallel}
\end{figure}

We should remark that the plane wave basis set provides a very efficient mean of calculating the electron--electron repulsion through the solution of the Poisson equation for the corresponding potential in Fourier space. By exploiting the fast Fourier transform (FFT), this problem is solved with an algorithm that scales as $\mathcal{O}(N\log N)$ with $N$ being the number of plane waves used to span the charge density. This is in contrast to localized orbital-based codes, where the Hartree potential is computed in real-space. This computation inherently scales non-linearly with the size of the atomic orbital basis set. The so-called ``freeze-and-thaw'' procedure \cite{weso1996b} by-passes this problem by evaluating the Hartree potential due to the frozen density only at the beginning of the calculation. This is made even more computationally amenable by employing density fitting techniques \cite{JCC:JCC20861}. 

We have, therefore, re-casted the FDE equations in such a way that the Hartree potential due to the total electron density can be calculated as often as each SCF cycle. This is achieved by imposing that the SCF cycle in Figure \ref{fig:SCF} is executed only once. In the implementation, this is achieved using Message-Passing Interface (MPI) as indicated in Figure \ref{fig:mpi}. For those interested in the details of the working equations, we include an appendix at the end of the paper.

\section{Computational Details}
All the calculations (KS-DFT and FDE) have been performed with {\sc Quantum ESPRESSO}. We have used Ultrasoft pesudopotentials from the original QE library and from \emph{GBRV}\cite{Garrity2014446} library. The plane wave cutoffs are \SI{40}{Ry} and \SI{400}{Ry}, for the wave functions and density, respectively. Unless stated otherwise, the PBE \cite{PhysRevLett.78.1396} functional has been employed for the exchange-correlation, and either the LC94 \cite{PhysRevA.50.5328} or revAPBEK \cite{prl106-186406} functionals have been employed for the non-additive KEDF.

Regarding sampling of the BZ, in the calculations invoving molecular systems, we have only sampled the $\Gamma$ point, while for periodic systems we have used the Monkhorst--Pack~\cite{monk1976} sampling with a $2\times 2\times 1$ mesh. In addition, we used a Methfessel--Paxton smearing~\cite{prb40-3616} with \SI{1e-2}{Ry} of smearing parameter. The smearing was only applied to the metallic subsystems, whereas the molecular subsystems were treated with integer orbital occupations.

We have benchmarked FDE calculations against KS-DFT references by comparing physisorption energies and corresponding electron densities obtained with the two methods. The simulations consist of single point calculations with the same geometry for both KS-DFT and FDE (e.g.\ the equilibrium geometry of the system calculated with KS-DFT with exception of the 2PDI on Au). Moreover, a more insightful comparison is made by calculating the number of electrons misplaced by FDE, $\braket{\Delta\rho}$, defined as:
\begin{equation}
 \braket{\Delta\rho} = \frac{1}{2} \int \left| \Delta\rho(\rvec) \right| \, d\rvec
\end{equation}
where
\begin{equation}
\label{eq:deltarho}
 \Delta\rho(\rvec) = \rho_\text{FDE}(\rvec) - \rho_\text{KS}(\rvec)
\end{equation}

$\braket{\Delta\rho}$ is an important quantity, as it vanishes only when FDE and KS-DFT electron densities coincide. As $\braket{\Delta\rho}$ is a size-sensitive quantity, we always compare it to the total number of subsystems and the total number of electrons.

\section{Results}
\label{sec:results}

We have chosen a set of molecular systems (water dimer and ammonia borane) as well as molecules on metal surfaces (\ce{CH4} on Pt, water on Pt and PDI on Au) as our set of pilot calculations. The goal of the calculations involving the molecular systems is to show that our FDE implementation reproduces the already reported behavior of FDE for these systems. Generally, FDE delivers results close to KS-DFT when the inter-subsystem density overlap is low. Thus, water dimer is expected to be well described by FDE, while for the ammonia borane system, due to the partial charge transfer and covalent character of the B--N dative bond, FDE is expected to fail. We will show that when tackling molecules adsorbed on surfaces, similar considerations to the molecular case apply. I.e.\ the larger the density in the region between subsystems, the less accurate FDE will be.

\begin{table}[p]
 \caption[Results summary]{Results summary Table. For each system studied, we have calculated the interaction energy, and the number of electrons that has been displaced by FDE with respect to the KS reference}
 \label{tab:results}
 \centering
 \begin{tabular}{lccccc}
  \toprule
   &\bf{Int. En. KS} & \bf{Int. En. LC94} & \bf{$\mathbf{\braket{\Delta\rho}}$ LC94} & \bf{Int. En. rAPBEK} & \bf{$\mathbf{\braket{\Delta\rho}}$ rAPBEK}\\
	 & (\SI{}{kcal/mol})	& (\SI{}{kcal/mol}) &	& (\SI{}{kcal/mol}) & \\
  \midrule
  \ce{H2O} Dimer &	$-4.317$ &	$-4.645$	&	$0.0285$ & $-4.472$ & $0.0291$ \\
  \ce{BH3-NH3} &	$-34.783$ &	$-80.865$	&	$0.1940$ & $-82.468$ & $0.1981$ \\
  \midrule
  \ce{CH4} on Pt T1 &	$-1.823$ &	$+1.302$	&	$0.0917$ & $+0.987$ & $0.0903$ \\
  \ce{CH4} on Pt T2 &	$-1.375$ &	$-0.492$	&	$0.0291$ & $-0.690$ & $0.0248$ \\
  \ce{CH4} on Pt T3 &	$-1.368$ &	$-0.493$	&	$0.0235$ & $-0.687$ & $0.0180$ \\
  \ce{CH4} on Pt B2a &	$-1.366$ &	$-0.433$	&	$0.0282$ & $-0.601$ & $0.0238$ \\
  \ce{CH4} on Pt B2p &	$-1.313$ &	$-0.606$	&	$0.0280$ & $-0.712$ & $0.0231$ \\
  \ce{CH4} on Pt H1 &	$-1.271$ &	$-0.546$	&	$0.0318$ & $-0.726$ & $0.0285$ \\
  \ce{CH4} on Pt H2 &	$-1.306$ &	$-0.540$	&	$0.0262$ & $-0.651$ & $0.0215$ \\
  \ce{CH4} on Pt H3 &	$-1.346$ &	$-0.535$	&	$0.0213$ & $-0.640$ & $0.0155$ \\
  \midrule
  \ce{H2O} on Pt (h) &	$-3.168$ &	$+0.040$	&	$0.1643$ & $+0.074$ & $0.1634$ \\
  \ce{H2O} on Pt (v) &	$-0.523$ &	$-0.162$	&	$0.0162$ & $+0.046$ & $0.0135$ \\
  \ce{12H2O} on Pt   &	$-135.470$ &	$-122.594$	&	$0.6487$ & $-117.824$ & $0.6421$ \\
  \midrule
  \ce{2PDI} on Au &	$+8.678$ &	$-17.078$	&	$0.3407$ & $-16.064$ & $0.3877$ \\
  \bottomrule
 \end{tabular}
\end{table}

\subsection{\ce{H2O} dimer}
\label{ssec:dimer}
\begin{figure}[p]
 \centering
 \subfloat[][]{\label{fig:Delta1-a}
 \includegraphics[width=.4\columnwidth]{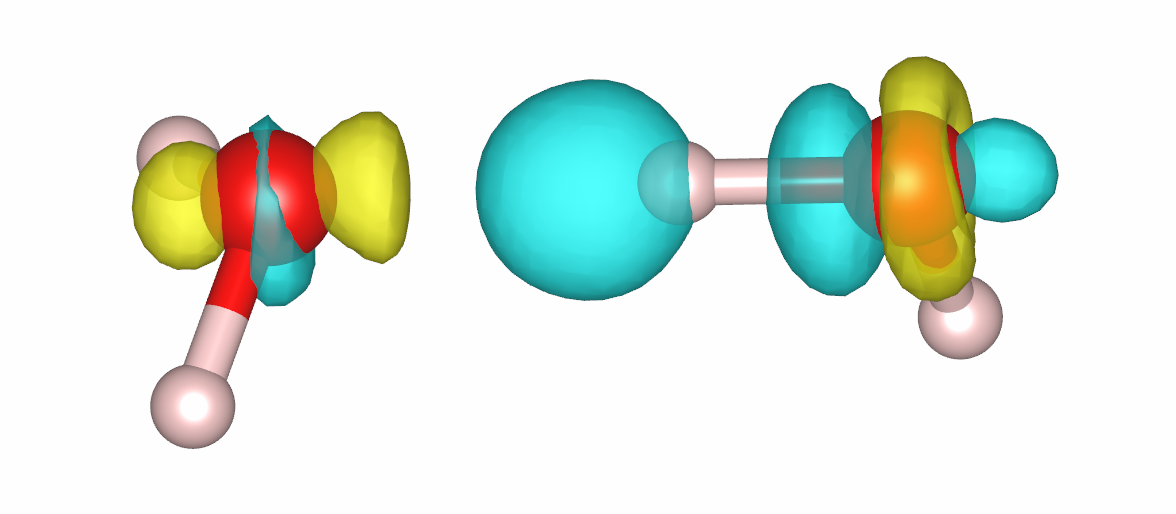}} \quad
 \subfloat[][]{\label{fig:Delta1-b}
 \includegraphics[width=.4\columnwidth]{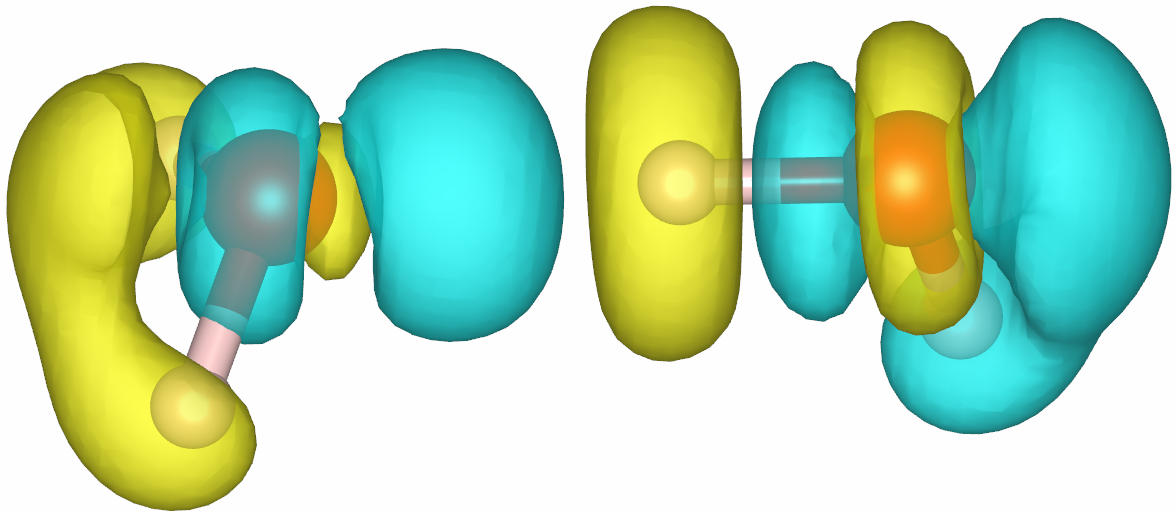}}
   \caption[Water dimer]{Water dimer. \subref{fig:Delta1-a}: Difference between the FDE density and the KS density; \subref{fig:Delta1-b}: Difference between the sum of the densities of isolated fragments and the KS density. In both cases a \num{1e-3} isosurface is plotted.}
 \label{fig:Delta1}
\end{figure}

The first system we used to benchmark our FDE code is the water dimer, with the two water molecules arranged to form a single hydrogen bond, see Figure \ref{fig:Delta1-a}.

It has been shown in previous publications~\cite{jcp106-8516,Iannuzzi200616} that localized basis set FDE is capable of obtaining very accurate results for this system when a GGA KEDF is employed for the non-additive kinetic energy part.
We have calculated an attractive interaction energy of \SI{4.317}{kcal} with regular KS-DFT, while FDE just slightly overestimates this energy by about \SI{0.3}{kcal} and in this particular case is well within what's considered chemical accuracy (\SI{1}{kcal}).

The very good agreement between between KS-DFT and FDE for the interaction energy is also reflected in the number of misplaced electrons $\braket{\Delta\rho}$. As reported in Table \ref{tab:results} only \num{2.85e-2} electrons out of 16 have been displaced by the selfconsistent FDE calculation. In Figure \ref{fig:Delta1-a} the difference between the FDE and KS-DFT densities, $\Delta\rho(\rvec)$, is plotted as defined in \equ{eq:deltarho}. The figure shows that the FDE method localizes too much density on the oxygen lone pairs in turn depleting  of electron density the hydrogen bond between the two dimers. It has to be pointed out that the \num{1e-3} isosurface is a very strict threshold. As a reference, Figure \ref{fig:Delta1-b} reports the difference between the KS-DFT density of the dimer and the sum of the density of the two isolated fragments. Using the same isosurface we can see that the difference is much larger than that of FDE. Thus, we conclude that for this system FDE performs very well, recovering almost exactly the same electronic structure obtained with a KS calculation.

\subsection{Ammonia Borane}
As shown previously \cite{cpl461-353}, the ammonia borane complex is expected to represent a challenge for FDE. It is a Lewis acid-base complex that exhibits a partial charge transfer from the ammonia lone pair to the borane, yielding a so-called dative bond characterized by a dissociation energy of \SI{34.783}{kcal/mol} as calculated by us using KS-DFT, in good agreement with coupled cluster calculations \cite{baus1995}. The FDE interaction energy of \SI{80.865}{kcal/mol} and \SI{82.468}{kcal/mol} significantly overestimates the reference when using LC94 or revAPBEK non-additive KEDF, respectively.

\begin{figure}[p]
 \centering
 \subfloat[][]{\label{fig:Delta4-a}
 \includegraphics[width=.4\columnwidth]{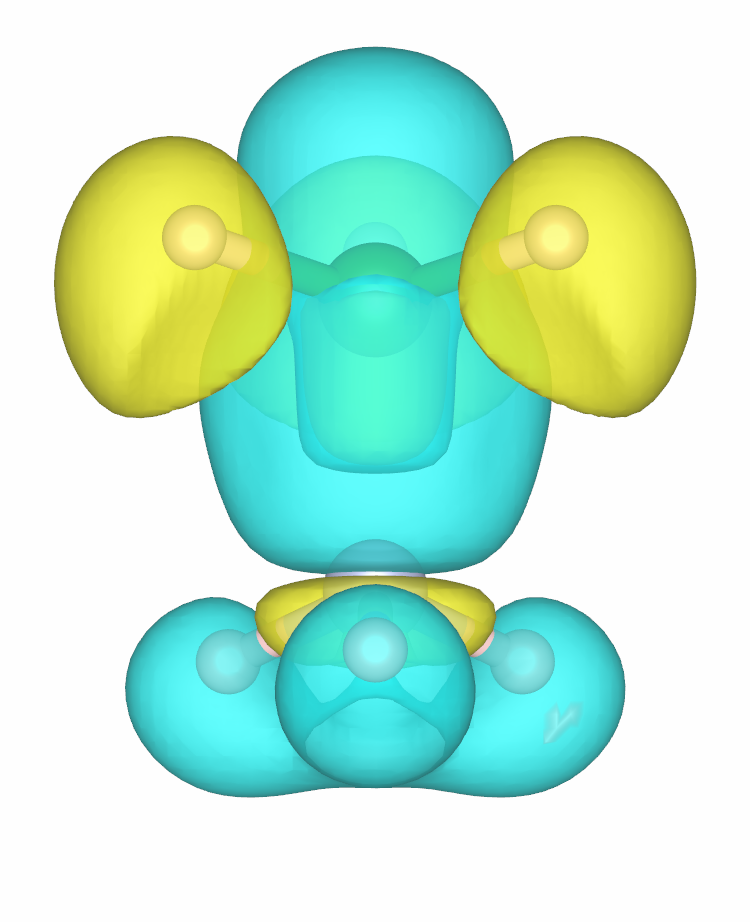}} \quad
 \subfloat[][]{\label{fig:Delta4-b}
 \includegraphics[width=.4\columnwidth]{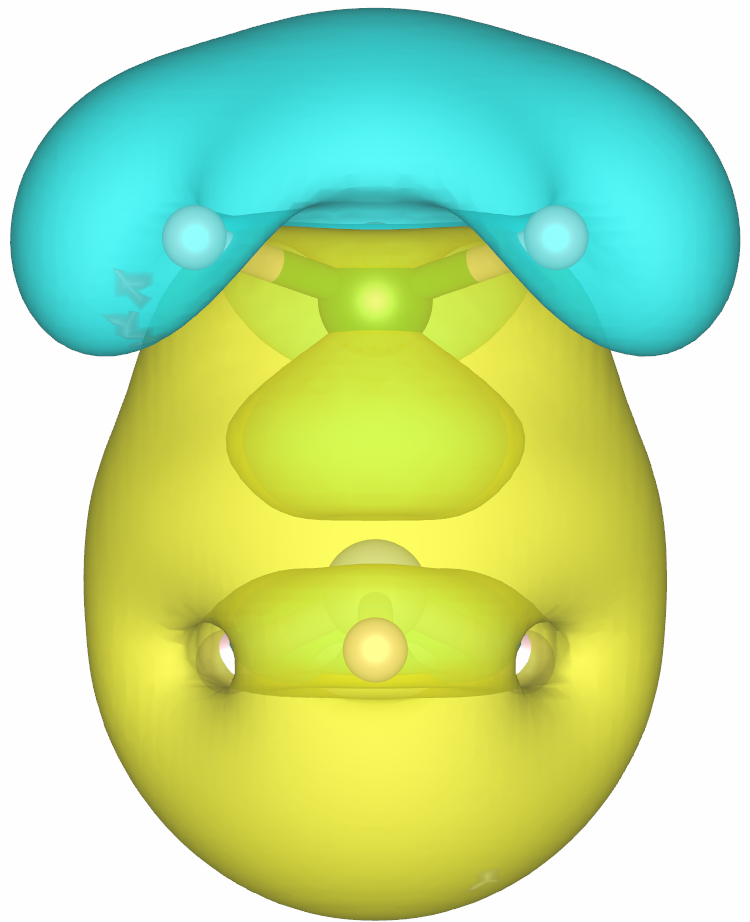}}
 \caption[Ammonia borane]{Ammonia borane complex. Borane on top. \subref{fig:Delta4-a}: Difference between the FDE density and the KS density; \subref{fig:Delta4-b}: Difference between the sum of the densities of isolated fragments and the KS density. In both cases a \num{1e-3} isosurface is plotted.}
 \label{fig:Delta4}
\end{figure}

More important is the analysis of the electron density: the difference between the FDE density (and superposition of the isolated fragments densities) with the supersystem KS density is reported in Figure~\ref{fig:Delta4-a} (Figure~\ref{fig:Delta4-b}). As the electron density undergoes significant changes upon formation of the complex, FDE only qualitatively reproduces the KS-DFT density --- there is a general underestimation of electron density along the B-N bond, while the hydrogens bonded to N (B) are too acidic (basic). Similar results are obtained from localized basis formulations of FDE \cite{cpl461-353}.

Quantitatively, the number of electrons misplaced by FDE in our calculations is \num{1.940e-1} and \num{1.981e-1} when employing the non-additive KEDF LC94 and revAPBEK, respectively. As expected, this is not satisfactory considering that the supersystem only has 14 electrons. In conclusion, we confirm previous FDE calculations of this complex \cite{cpl461-353} and we also show that our plane wave implementation maintains quantitatively the FDE electronic structure predicted by atomic orbital based implementations. 

\subsection{\ce{CH4} on Pt(100)}
Catalitic activation of methane by adsorption on a transition metal surface represents an intresting problem, as it could represent a cheap way to produce hydrogen, or a precursor in the synthesis of more complex molecules.

\begin{figure}[p]
 \centering
 \subfloat[][T1]{\label{fig:CH4-a}
 \includegraphics[width=.125\columnwidth]{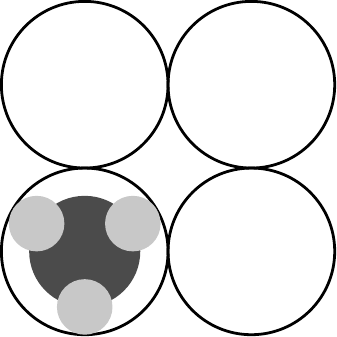}}
 \subfloat[][T2]{\label{fig:CH4-b}
 \includegraphics[width=.125\columnwidth]{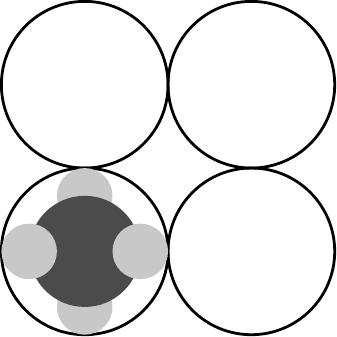}}
 \subfloat[][T3]{\label{fig:CH4-c}
 \includegraphics[width=.125\columnwidth]{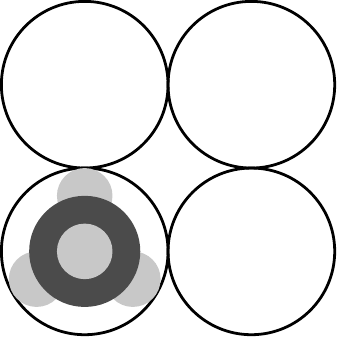}}
 \subfloat[][B2a]{\label{fig:CH4-d}
 \includegraphics[width=.125\columnwidth]{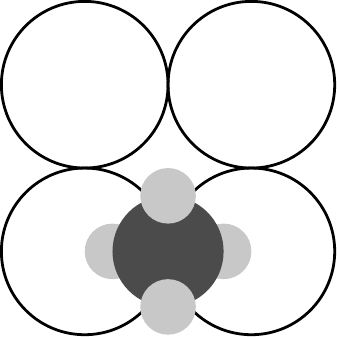}} 
 \subfloat[][B2p]{\label{fig:CH4-e}
 \includegraphics[width=.125\columnwidth]{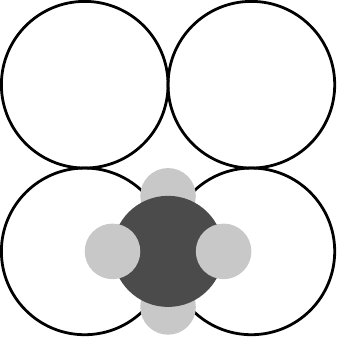}}
 \subfloat[][H1]{\label{fig:CH4-f}
 \includegraphics[width=.125\columnwidth]{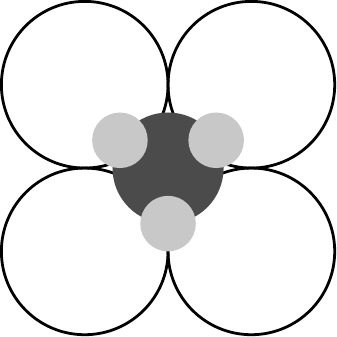}}
 \subfloat[][H2]{\label{fig:CH4-g}
 \includegraphics[width=.125\columnwidth]{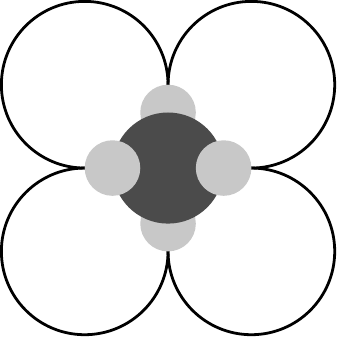}}
 \subfloat[][H3]{\label{fig:CH4-h}
 \includegraphics[width=.125\columnwidth]{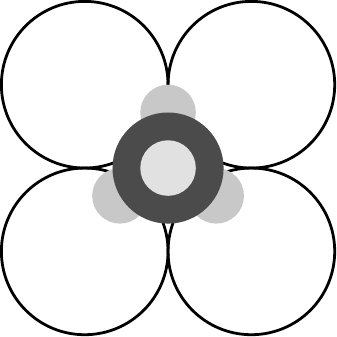}}
 \caption[CH$_4$ Conformations]{Configurations of methane adsorbed on Pt(100) considered in the calculations.}
 \label{fig:CH4}
\end{figure}

On a \ce{Pt(100)} surface methane can be adsorbet on several sites, so-called top, bridge, and hollow. In addition the molecule can have several configurations with respect to the surface. We have tested the behavior of FDE against the 8 configurations shown in figure~\ref{fig:CH4}.

As it has been obeserved in Ref.\ \cite{ss549-231} and confirmed in our calculations (Table~\ref{tab:results}) the \ce{CH4-Pt} interaction is generally very weak, and the potential energy surface is relatively uncorrugated (i.e. insensitive to the methane orientation). The only clear potential well is represented by the T1 configurations, which has an interaction energy about \SI{0.5}{kcal/mol} higher than all the others. From an analysis of the local density of states of this configuration, it has been concluded that in the T1 case we are dealing with a real chemisorption (electron backdonation from the the metal \emph{d} orbitals to the carbon), rather than a physisorption (polarization of the metal-methan electron densities).

We can see that FDE reproduces the supersystem KS calculation in a very accurate way for all configurations but the T1: interaction energies are always within chemical accuracy, and the number of displaced electrons is very small, ranging from \num{2.1e-2} to \num{3.2e-2} for the LC94 functional, and from \num{1.5e-2} to \num{2.5e-2} for the revAPBEK functional. It has to be remarked that the revAPBEK functional provided the best results also for magnitude of the interaction energy.

\begin{figure}[p]
 \centering
 \includegraphics[width=.4\columnwidth]{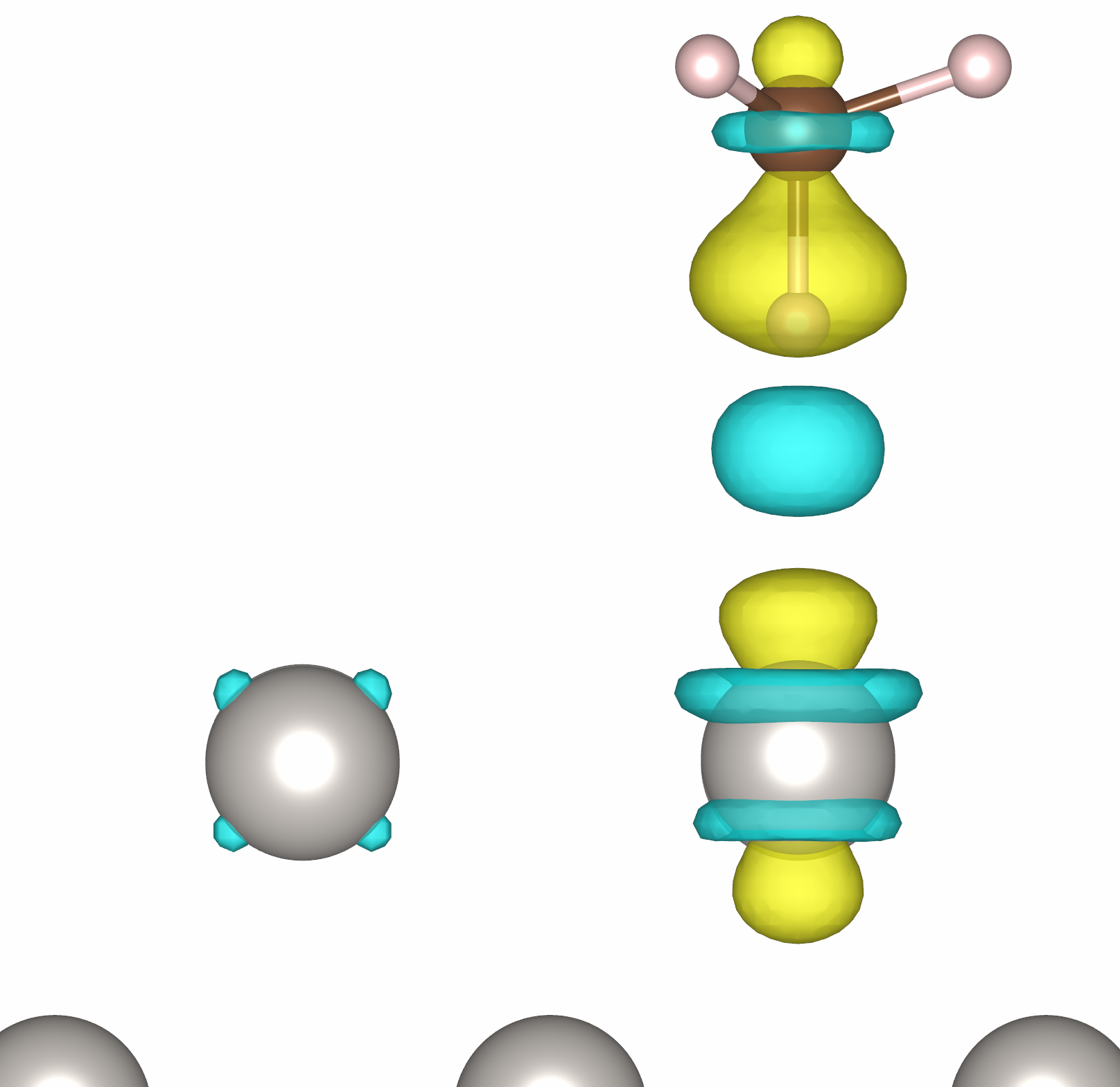}
 \caption[]{Methane on Pt $\Delta\rho(\rvec)$ 0.001 isosurface plot. Yellow identifies positive regions (excess of FDE density), blue identifies negative regions (excess of KS density).}
 \label{fig:Delta3}
\end{figure}

As expected the T1 configuration is the one for which FDE yielded the worst results, because of the chemical nature of the bond between the hydrocarbon and the surface: we can see that FDE predicts a small repulsive interaction, equal to \SI{1.302}{kcal/mol} for LC94 and \SI{0.987}{kcal/mol} for revAPBEK, whereas KS has an attractive energy of \SI{1.823}{kcal/mol}. The number of electrons misplaced $\Delta\rho$ is about \num{9e-2}, which is which is much higher than what abserved for all the other configurations, but still reasonably small. In Figure~\ref{fig:Delta3}, an isosurface plot of $\Delta\rho(\rvec)$ for the T1 configuration is reported. Again, we see that the selfconsistent FDE density is smaller than the KS-DFT one in the interfragment bonding region.

\subsection{\ce{H2O} on Pt(111)}
The most complex system we have used to benchmark this periodic implementation of FDE is a water bilayer adsorbed on a \ce{Pt(111)} surface. The system consists of 13 fragments, 12 water molecules plus the metal surface. The geometry we have employed to model the water bilayer is the so-called \emph{RT3}, as described in Refs.\ \cite{Doering1982305,PhysRevB.69.195404}

On the layer closer to the surface, we can distinguish between two kinds of water molecules: the ones with the dipole moment parallel to the metal surface, and the ones with a non-zero dipole component in the perpendicularly to the surface. The intermolecular interactions playing an important role in keeping the the bilayer and the metal together are three: (1) the hydrogen bonds network, (2) the interaction between the surface and the parallel dipole water molecules, and (3) the interaction between the surface and the water molecules having non-zero perpendicular component of the dipole.

We have already seen in Section~\ref{ssec:dimer} that FDE describes well hydrogen bonds. This system, however, will probe the ability of FDE to describe the other two cases. Thus, we have run two additional simulations:
\begin{enumerate}
 \item A single parallel water molecule adsorbed on the surface;
 \item A single non-parallel water molecule adsorbed on the surface;
\end{enumerate}
In both cases the position of the molecule is the same as that in the bilayer, without performing any additional optimization. We will describe these two additional calculations and the one pertaining the full bi-layer separately below.

\subsubsection{Parallel \ce{H2O}}
For the parallel water molecule, KS predicts an adsorption energy of \SI{3.168}{kcal} while according to FDE the molecule is unbound, with and adsorption energy very close to zero, see Table~\ref{tab:results}.

This discrepancy is also reflected in the number of misplaced electrons $\Delta\rho$, that for this system is the highest so far and equal to \num{1.643e-1}. Plots of $\Delta\rho(\rvec)$ in Figure~\ref{fig:Delta2-a} show that once again FDE places too much density on the oxygen lone pairs and not enough in the bonding region between the molecule and the surface.

\subsubsection{Perpendicular \ce{H2O}}
The vertical water molecules are further from the surface, and interact with it very weakly, \SI{0.523}{kcal} as predicted by KS, and \SI{0.162}{kcal} according to FDE. Being the density of the two fragments almost non-overlapping, obviously FDE has no problems recovering the same total density as KS, with only \num{1.62e-2} electrons displaced.

\subsubsection{Water Bilayer}
As we have seen in the previous sections, we expect that for the bilayer the largest source of error is due to the water molecules in the first layer parallel to the surface and  to a lesser extent due to the hydrogen bond network.
\begin{figure}[p]
 \centering
 \subfloat[][]{\label{fig:Delta2-a}
 \includegraphics[width=.4\columnwidth]{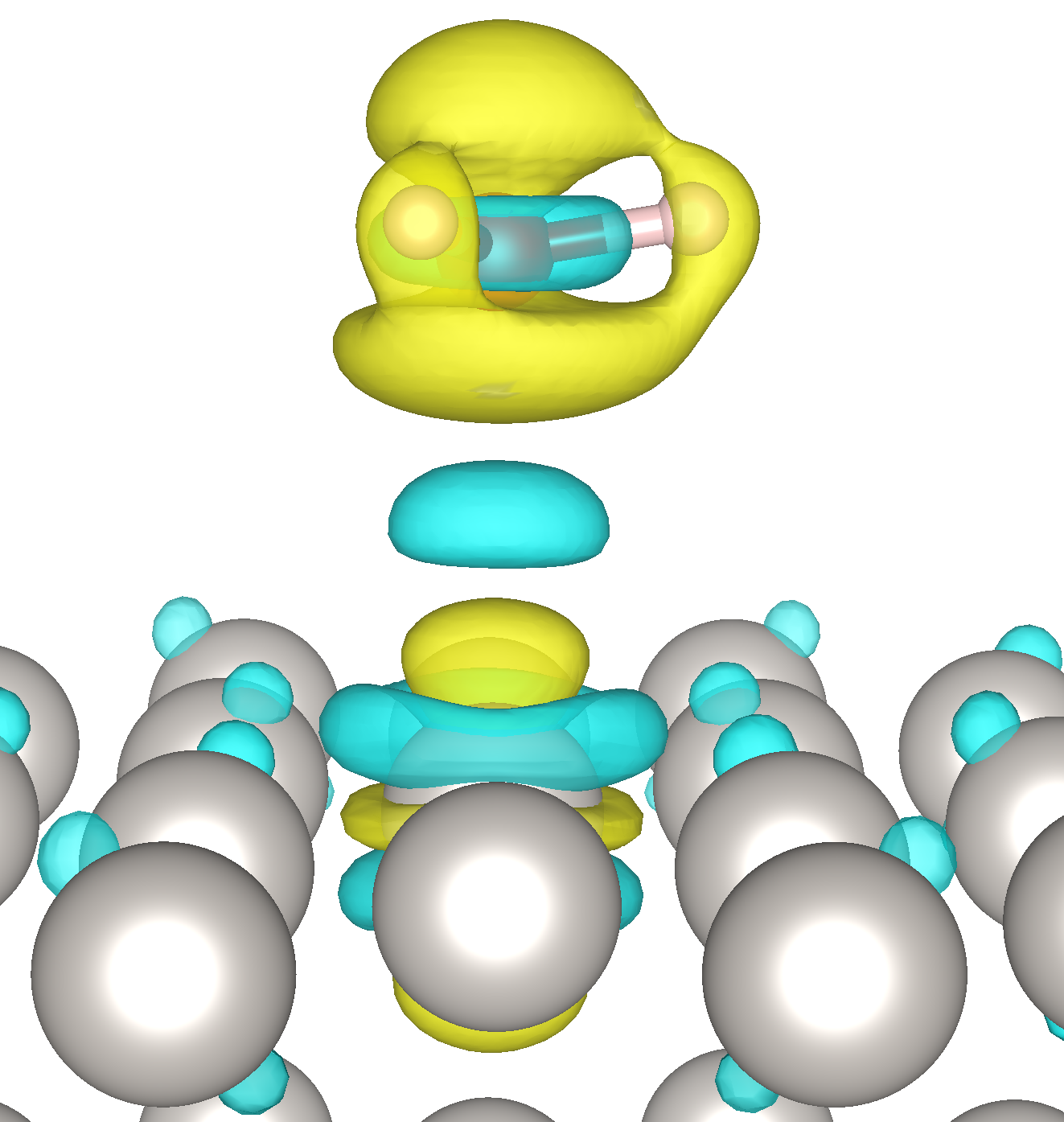}} \quad
 \subfloat[][]{\label{fig:Delta2-b}
 \includegraphics[width=.4\columnwidth]{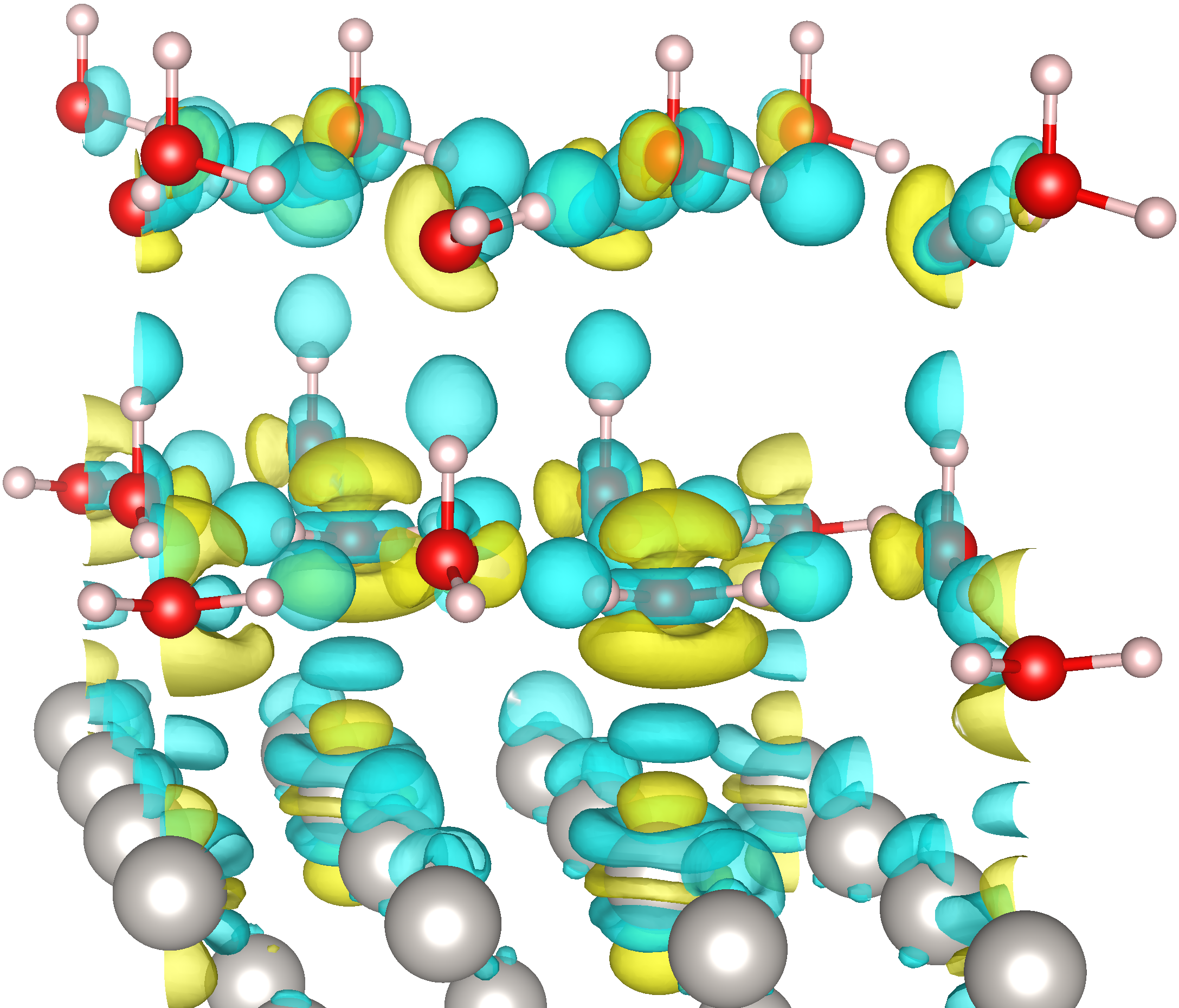}}
 \caption[Water on Pt]{Water on Pt \subref{fig:Delta2-a} and water bilayer on Pt \subref{fig:Delta2-b} $\Delta\rho(\rvec)$ 0.001 isosurface plot. Yellow identifies positive regions (excess of FDE density), blue identifies negative regions (excess of KS density).}
 \label{fig:Delta2}
\end{figure}

The total intermolecular interactions as predicted by KS-DFT is \SI{135.147}{kcal}, while FDE underestimates it by just 10\% yielding an interaction equal to \SI{122.594}{kcal}. As shown in Table~\ref{tab:results} only \num{6.487e-1} out of the total 368 are misplaced by FDE in this complex 13 fragments system. An isosurface plot of $\Delta\rho(\rvec)$ is presented in Figure~\ref{fig:Delta2-b}. In the figure we see that the expected maximum discrepancy between the KS-DFT and FDE densities is in the bond between the parallel water molecules and the surface, as well as in the many hydrogen bonds keeping the bilayer together.

\subsection{Perylene Diimide on Au(111)}

Perylene diimides (PDIs) are a class of compounds whose electron-accepting character and charge transport properties make them useful materials in the design of new photovoltaic cells based on organic compounds.

We have run a single point KS simulation of two N,N'-dihydro perylene diimides stacked on top of a Au surface. The geometry has been optimized at the KS-DFT level constraining the PDI--PDI distance and the PDI--Au distance to be 3 \AA.

From the KS-DFT calculation of the supersystem, it has been predicted a \SI{8.678}{kcal/mol} repulsive interaction energy. The positive sign in the interaction energy is not surprising since van der Waals (vdW) forces are the most important intermolecular forces in this case, and it is known that KS-DFT calculations employing semilocal GGA exchange--correlation functionals is generally unable to recover them.

Conversely, FDE predicts an attractive interaction energy of \SI{17.078}{kcal/mol} and \SI{16.064}{kcal/mol} with LC94 and revAPBEK, respectively. Despite not reproducing the KS-DFT interaction energy, this result is correctly attractive. Such a behavior from FDE has been reported previously \cite{kevo2006,kevo2014b}.

As reported in Table~\ref{tab:results}, the number of displaced electrons is \num{3.4e-1} with LC94 and \num{3.9e-1} with revAPBEK -- a negligible value considering that each PDI molecule has 140 electrons.

\section{Conclusions}
In this work, we have extended the Frozen Density Embedding formulation of subsystem DFT to treat inherently periodic subsystems. In FDE, the subsystems are treated as isolated Kohn--Sham systems which interact through an effective embedding potential. This potential is a functional of the electron density of all subsystems. We introduce periodic boundary conditions by applying the Bloch theorem to the Kohn--Sham orbitals of the subsystems. Thus, sampling over the Brillouin zone (the so-called $k$-point sampling) enters the working Kohn--Sham-like equations needed to recover the electronic structure (electron density) of each subsystem. 

The ultimate goal of this approach is to reduce the computational complexity of KS-DFT for systems composed of molecules interacting with surfaces of metals and semiconductors. Several open questions have been answered here. When semilocal GGA kinetic energy functionals are employed, our implementation of FDE reproduces the results of existing molecular implementations (i.e.\ implementations employing a localized orbital basis) for interacting molecules. We have established this connection for those systems which are well described by FDE and for those which semilocal FDE is known to fail (such as the ammonia borane complex). We have also run pilot calculations to establish how well FDE tackles weakly interacting molecules at metal surfaces. We have established that FDE yields chemical accuracy when, in the KS-DFT calculation, there is no hybridization of the orbitals of the molecules with the bands of the metal. 

With such encouraging results, this work sets the stage to applying FDE to a wide array of molecular systems interacting with surfaces with applications to energy-related materials as well as catalysis and photovoltaics. 

Because of the inapplicability of semilocal FDE to molecule--surface interactions beyond weak physisorption, we also conclude that non-locality in the energy functionals (both at the kinetic and the exchange--correlation level) must be included if FDE aspires to become a true alternative to Kohn--Sham DFT. Efforts in this directions have been taken in our lab regarding the exchange--correlation \cite{kevo2014b}, and ongoing efforts are placed in the search for computationally amenable ways to include non-locality at the kinetic energy level as well.

\section{Acknowledgements}
We thank Dr.\ Oliviero Andreussi for stimulating discussions. We acknowledge start-up funds provided by Rutgers University--Newark, for support of this research.

\section{Appendix}

\subsection{Recast of the Effective Potential expression}

Due to the fact that in molecular codes employing localized orbital basis sets the computation of two-electron integrals can be expensive (Gaussian basis), or only possible through density fitting (Slater basis), the Hartree energy and potential in the subsystem DFT framework is usually computed as:
\begin{equation}
 E_H[\rho] = \frac{1}{2} \sum_I^{N_S} \left[ \int V_H[\rho_I](\rvec)\rho_I(\rvec) \, d\rvec + \int V_H^{I\text{emb}}[\rho_{J \neq I}](\rvec)\rho_I(\rvec) \, d\rvec \right],
\end{equation}
where
\begin{gather}
 V_H[\rho_I](\rvec) = \int \frac{\rho_I(\rvec')}{|\rvec - \rvec'|} \, d\rvec' \\
 V_H^{I\text{emb}}[\rho_{J \neq I}](\rvec) = \sum_{J\neq I}^{N_S} \int \frac{\rho_J(\rvec')}{|\rvec - \rvec'|} \, d\rvec' \equiv \sum_{J\neq I}^{N_S} V_H[\rho_J](\rvec) \equiv V[\rho](\rvec) - V[\rho_I](\rvec),
\end{gather}
from which is clear that we are indeed calculating the total Hartree energy in a similar way as the kinetic energy:
\begin{equation}
 E_H[\rho] = \sum_I^{N_S} E_H[\rho_I] + E_H[\rho] - \sum_I^{N_S} E_H[\rho_I] \equiv \sum_I^{N_S} E_H[\rho_I] + E_H^\text{nad}.
\end{equation}
This approach is useful for the freeze-and-thaw scheme implemented in molecular codes. In this scheme, $V_H^{I\text{emb}}(\rvec)$ is calculated once and it doesn't change until the end of the SCF cycle of the active fragment.

The plane wave basis set allows us to efficiently calculate the potential due to the total electron density, $V_H[\rho]$. Thus, we have the possibility to do so on the fly at each SCF step (and we almost always do). In this way, there is no need to calculate the embedding potential in \equ{eq:embedding}, and the effective potential appearing in the KS-like equations, \equ{eq:KS}, becomes:
\begin{equation}
 V_\text{eff}^I[\rho,\rho_I](\rvec) = V_{eN}(\rvec) + V_H[\rho](\rvec) + V_{XC}[\rho_I](\rvec) + V_\text{kin}^\text{nad}[\rho,\rho_I](\rvec) + V_\text{XC}^\text{nad}[\rho,\rho_I](\rvec)
\end{equation}
where in $V_{eN}(\rvec)$ we have grouped the interactions with all nuclei in the system.
\subsection{Recast of the Total energy expression}
To better exploit the {\sc Quantum ESPRESSO} code base, we found useful to express the FDE energy as a sum of band energies (i.e. the sum of eigenvalues of the occupied states), and a sum of the energy of the single fragments.

For a single fragment the band energy is:
\begin{equation}\begin{split}
\label{eq:eband}
 \sum_k \epsilon_k^I &= \sum_k \braket{\psi_k^I| - \frac{1}{2} \nabla^2 + V_\text{eff}^I[\rho,\rho_I] |\psi_k^I} \\
			     &= T_J[\rho_I] + \int V_{eN}(\rvec) \, \rho_I(\rvec) \, d\rvec + \int V_H[\rho](\rvec) \, \rho_I(\rvec) \, d\rvec \\
			     & \qquad \qquad + \int V_{XC}[\rho_I](\rvec) \, \rho_I(\rvec) \, d\rvec + \int \tilde{V}_{XC}[\rho](\rvec) \, \rho_I(\rvec) \, d\rvec \\
			     & \qquad \qquad - \int \tilde{V}_{XC}[\rho_I](\rvec) \, \rho_I(\rvec) \, d\rvec + \int \tilde{V}_\text{kin}[\rho](\rvec) \, \rho_I(\rvec) \, d\rvec \\
			     & \qquad \qquad - \int \tilde{V}_\text{kin}[\rho_I](\rvec) \, \rho_I(\rvec) \, d\rvec
\end{split}\end{equation}

If we define the quantity $V_K\rho_K$ (found in the code as the \verb|deband| variable) as
\begin{equation}
 V_K\rho_K = - \int \left( V_H[\rho] + V_{XC}[\rho_K] + \tilde{V}_{XC}[\rho] - \tilde{V}_{XC}[\rho_K] + \tilde{V}_\text{kin}[\rho] - \tilde{V}_\text{kin}[\rho_K] \right) \, \rho_K(\rvec) \, d\rvec
\end{equation}
we see from \equ{eq:eband} that:
\begin{equation}
 \sum_i^{n_K} \epsilon_{i_K} + V_K\rho_K = T_s[\rho_K] + E_{eN} [\rho_K]
\end{equation}

And defining the fragment energy as:
\begin{equation}\begin{split}
 E_K &= \sum_i^{n_K}  \epsilon_{i_K} + V_K\rho_K + E_{XC}[\rho_K] + E_H[\rho_K] + V_{NN} \\
     &\equiv T_s[\rho_K] + E_{eN} [\rho_K] + E_{XC}[\rho_K] + E_H[\rho_K] + V_{NN}
\end{split}\end{equation}
we finally obtain that the FDE energy is simply:
\begin{equation}
  E_\text{FDE} = \sum_K^N E_K + T_s^\text{nad} + E_{XC}^\text{nad} - (N-1)V_{NN} 
\end{equation}

\subsection{Kinetic energy functional and potential}
We have implemented an array of semilocal kinetic energy functional in our code. The general form of these functionals is:
\begin{equation}
 T_s^\text{nad}[\rho,\nabla\rho] = C_{TF} \int dr\, \rho^{5/3}(r) F(s(r))
\end{equation}
where
\begin{gather}
  s(\rvec) = C_s \frac{|\nabla\rho(\rvec)|}{\rho^{4/3}(\rvec)} \\
  C_{TF} = \frac{3}{10} (3\pi^2)^{2/3} \\
  C_s = \frac{1}{2(2\pi^2)^{1/3}}
\end{gather}

For a generic functional
\begin{equation}
 F[\rho,\nabla\rho] = \int dr\, G(\rho(r),\nabla\rho(r))
\end{equation}
its functional derivative is given by
\begin{equation}
   \frac{\delta F}{\delta\rho(\rvec)} = \left.\frac{\partial G}{\partial \rho}\right\vert_{\rho=\rho(\rvec)} - \left.\nabla \cdot
    \frac{\partial G}{\partial \nabla\rho}\right\vert_{\rho=\rho(\rvec)}
\end{equation}

We can therefore calculate the kinetic energy potential for any of these functionals
\begin{gather}
 \frac{\partial G}{\partial\rho} = \frac{5}{3}\rho^{2/3}F(s) + \rho^{5/3}
    \frac{\partial F}{\partial s}\frac{\partial s}{\partial\rho} = 
    \rho^{2/3} \left[ \frac{5}{3}F(s) - \frac{4}{3}s(r)\frac{\partial F}{\partial s}
    \right] \\
  \frac{\partial G}{\partial\nabla\rho} = \rho^{5/3}\frac{\partial F}{\partial s}
    \frac{\partial s}{\partial|\nabla\rho|}\frac{\nabla\rho}{|\nabla\rho|} =
    C_s \rho^{1/3} \frac{\partial F}{\partial s} \frac{\nabla\rho}{|\nabla\rho|}
\end{gather}

Therefore
\begin{equation}\begin{split}
 \der{T_s^\text{nad}[\rho,\nabla\rho]}{\rho} &= C_{TF} \left\{  
	 \rho^{2/3} \left[ \frac{5}{3}F(s) - \frac{4}{3}s(r)\frac{\partial F}{\partial s} \right] 
	 -\nabla \cdot \left[ C_s \rho^{1/3} \frac{\partial F}{\partial s} \frac{\nabla\rho}{|\nabla\rho|} \right] \right\} \\
	 &\equiv \tilde{V}_\text{kin}[\rho](\rvec)
\end{split}\end{equation}

\subsection{Handling of non-local pseudopotentials}
In all {\sc Quantum ESPRESSO} calculations carried out in this work, either Ultrasoft (US)~\cite{USPP} or Projector Augmented Wave (PAW)~\cite{PAW} pseudopotentials are employed.

In the PAW formalism the true all-electron single particle wavefunction $\psi_n$ can be recovered from a smooth auxiliary wavefunction $\tilde{\psi}_n$ through the transformation
\begin{equation}
\label{eq:PAW1}
 \ket{\psi_n} = \hat T \ket{\tilde\psi_n}
\end{equation}
where the short label $n$ identifies a $\vec{k}$, band, and spin index.

The $\hat T$ operator can be written as:
\begin{equation}
\hat T
        = \mathbf{1} + \sum_a \sum_i \left(\ket{\phi_i^a} - \ket{\tilde \phi_i^a}\right)\bra{\tilde p_i^a}
\end{equation}
where $a$ labels the augmentation sphere of the ions (i.e.\ the nuclei), the so-called partial waves $\ket{\phi_i^a}$ form a complete basis set, and the auxiliary $\ket{\tilde{\phi}_i^a}$ obey $\ket{\phi_i^a}= \hat T \ket{\tilde\phi_i^a}$, and the set of the \emph{smooth projectior functions} $\bra{\tilde p_i^a}$ has the following properties:
\begin{gather}
    \braket{\tilde p_i^a | \tilde \phi_j^a} = \delta_{ij} \\
    \sum_i \ket{\tilde \phi_i^a}\bra{\tilde p_i^a}  = \mathbf{1} \\
    \tilde p_i^a (\rvec) \equiv \braket{\rvec | \tilde p_i^a} = 0 , \quad \text{for} \quad |\rvec - \vec{R}^a| > r_c^a
\end{gather}

In the \emph{frozen core approximation} of PAW the core electron are considered as frozen, and the eigen value problem is restricted only to the valence electrons. The full valence wave function obtained from \equ{eq:PAW1} is guaranteed to be orthogonal with respect to the core states.

Since in the FDE formalism the KS-DFT orbitals of one subsystem are not required to be orhogonal to those (valence nor core) in another subsystem (the non-additive kinetic energy potential is responsible of recovering those properties that in the KS method are guaranteed from orthonormality), the operator $\hat T$ describing the core states of one subsystem does not need to include projectors belonging to other subsystems. Thus, a subsystem-specific PAW projector must be used. Namely,
\begin{equation}
\hat T^I
        = \mathbf{1} + \sum_{a\in I} \sum_i \left(\ket{\phi_i^a} - \ket{\tilde \phi_i^a}\right)\bra{\tilde p_i^a}.
\end{equation}

%
%


\end{document}